\documentclass{ws-ijgmmp} 
\usepackage{amsmath}
\usepackage{amssymb}
\numberwithin{equation}{section}

\begin{document}
\markboth{G. Esposito}
{A parametrix for quantum gravity?}

\title{A PARAMETRIX FOR QUANTUM GRAVITY?}

\author{GIAMPIERO ESPOSITO}
\address{Istituto Nazionale di Fisica Nucleare, Sezione di
Napoli, Complesso Universitario di Monte S. Angelo, 
Via Cintia Edificio 6, 80126 Napoli, Italy\\
\email{gesposit@na.infn.it}}

\maketitle

\begin{abstract}
In the sixties, DeWitt discovered that the advanced and retarded Green functions of the wave operator on
metric perturbations in the de Donder gauge make it possible to define classical Poisson brackets on the
space of functionals that are invariant under the action of the full diffeomorphism group of spacetime.
He therefore tried to exploit this property to define invariant commutators for the quantized gravitational
field, but the operator counterpart of such classical Poisson brackets turned out to be a hard task. On the
other hand, in the mathematical literature, it is by now clear that, rather than inverting exactly an 
hyperbolic (or elliptic) operator, it is more convenient to build a quasi-inverse, i.e. an inverse operator
up to an operator of lower order which plays the role of regularizing operator. This approximate inverse, 
the parametrix, which is, strictly, a distribution, 
makes it possible to solve inhomogeneous hyperbolic (or elliptic) equations. We here suggest
that such a construction might be exploited in canonical quantum gravity provided one understands what is the
counterpart of classical smoothing operators in the quantization procedure. We begin with the simplest case,
i.e. fundamental solution and parametrix for the linear, scalar wave operator; the next step are tensor wave
equations, again for linear theory, e.g. Maxwell theory in curved spacetime. 
Last, the nonlinear Einstein equations are studied,
relying upon the well-established Choquet-Bruhat construction, according to which the fifth derivatives of 
solutions of a nonlinear hyperbolic system solve a linear hyperbolic system. The latter is solved by
means of Kirchhoff-type formulas, while the former fifth-order equations can be solved by means of
well-established parametrix techniques for elliptic operators. But then the metric components that solve
the vacuum Einstein equations can be obtained by convolution of such a parametrix with 
Kirchhoff-type formulas. Some basic functional equations for the parametrix are also obtained, that help
in studying classical and quantum version of the Jacobi identity. 
\end{abstract}


\section{Introduction}

The Hamiltonian road to quantization has played a key role, over the last century, in the development of quantum
mechanics \cite{Dirac1926}, quantum field theory in flat spacetime \cite{IZ1985}, 
including quantum Yang-Mills \cite{YM2005} and the particle physics
standard model, as well as in the formulation of canonical quantum gravity 
\cite{Dirac1958,DW1967a,Wheeler,Ashtekar,Rovelli,Thiemann,RV2015}. The main drawback of the Hamiltonian formulation, 
despite its beautiful and powerful applications to the classical Cauchy problem of general relativity
\cite{Foures1952,ADM,Choquet2009}, lies in the loss of the full diffeomorphism group of four-dimensional
spacetime, with the associated undoing of the unification of space and time into the spacetime manifold
(see, however, the valuable work in Refs. \cite{IK85a,IK85b} on the way to circumvent this problem).

Indeed, at classical level, the tools of global differential geometry make it possible to obtain a spacetime
covariant formulation of the constraint equations \cite{EGS1995}, which turn out to be linear \cite{ES1996}
on a bigger space, the space of multimomenta. Even earlier, at quantum level, the work of Peierls
\cite{Peierls1952} and DeWitt \cite{DeWitt1960} made it possible to define a Poisson bracket on the space of
all field functionals that remain invariant under the action of the infinite-dimensional Lie
(pseudo-)group of the theory (for gravity, this is the group of spacetime diffeomorphisms). If $A$ and $B$
are any two such functionals of the field variables $\varphi^{i}(x)$, their classical Peierls bracket reads
as (see Appendix for the notation)
\begin{equation}
(A,B) \equiv A_{,i} {\widetilde G}^{ij}B_{,j}
=\int {\rm d}^{4}x \int {\rm d}^{4}z \; 
{\delta A \over \delta \varphi^{i}}(x) {\widetilde G}^{ij}(x,z)
{\delta B \over \delta \varphi^{j}}(z), 
\label{(1.1)}
\end{equation}
where ${\widetilde G}^{ij}$, the supercommutator function \cite{DW1984,DW2003}, is the difference between
advanced and retarded Green functions for the invertible operator $F_{ij}$ acting on fields:
\begin{equation}
{\widetilde G}^{ij} \equiv G^{+ij}-G^{-ij}=-{\widetilde G}^{ji}.
\label{(1.2)}
\end{equation}
The advanced and retarded Green functions are both left- and right-inverses of $F_{ij}$, i.e.
\begin{equation}
G^{\pm ij'}F_{j'k''}=-\delta_{\; k''}^{i}=-\delta_{k}^{i} \delta(x,x''),
\label{(1.3)}
\end{equation}
\begin{equation}
F_{ij'}G^{\pm j'k''}=-\delta_{i}^{\; k''}=-\delta_{i}^{k}\delta(x,x'').
\label{(1.4)}
\end{equation}

In the framework of Ref. \cite{DeWitt1960}, the state of the art on the application of such ideas
to Einstein's gravity was as follows.
\vskip 0.3cm
\noindent
(i) The absolute invariants (also called observables or gauge-invariant functionals) $A$ and $B$ satisfy, by
definition, the conditions
\begin{equation}
\nabla_{\nu}{\delta T \over \delta g_{\mu \nu}}=0, \; T=A,B,
\label{(1.5)}
\end{equation}
which are a particular case of the gauge-invariance equation $Q_{\alpha}T=0$ of Appendix A (see (A5)),
with $T=A,B$, because the infinitesimal diffeomorphism 
\begin{equation}
\delta g_{\mu \nu}=\nabla_{\nu}\xi_{\mu}+\nabla_{\mu}\xi_{\nu}
\label{(1.6)}
\end{equation}
is a particular case of the general equation 
$\delta \varphi^{i}=Q_{\; \alpha}^{i}\delta \xi^{\alpha}$ of Appendix A (see (A3)).
\vskip 0.3cm
\noindent
(ii) The infinitesimal variation suffered from $A$, say, reads as
\begin{equation}
\delta^{\pm}A=\int {\delta A \over \delta g_{\mu \nu}}\delta^{\pm}g_{\mu \nu}{\rm d}^{4}x.
\label{(1.7)}
\end{equation}
Since the spacetime metric is not an invariant, its variations $\delta^{\pm}g_{\mu \nu}$ are determined
only up to a coordinate transformation (see (1.6)), while Eqs. (1.5) guarantee that the advanced and 
retarded variations $\delta^{\pm}A$ do not suffer from redundancies.
\vskip 0.3cm
\noindent
(iii) If $S$ is the Einstein-Hilbert action, the associated Euler-Lagrange equations are
(hereafter $g \equiv -{\rm det}g_{\mu \nu}$)
\begin{equation}
0={\delta S \over \delta g_{\mu \nu}}=G^{\mu \nu}=\sqrt{g}\left(R^{\mu \nu}
-{1 \over 2}g^{\mu \nu}R \right).
\label{(1.8)}
\end{equation}
If the action functional undergoes the change
\begin{equation}
S \rightarrow S + \varepsilon B,
\label{(1.9)}
\end{equation}
with constant parameter $\varepsilon$, the new Euler-Lagrange equations read as
\begin{equation}
\delta^{\pm}G^{\mu \nu}=-\varepsilon {\delta B \over \delta g_{\mu \nu}}.
\label{(1.10)}
\end{equation}
The general solution of Eq. (1.10) is obtained by adding (1.6) to a particular solution determined by
appropriate boundary and supplementary conditions. A convenient form of supplementary (or 
gauge-fixing) condition is \cite{DeWitt1960}
\begin{equation}
\left(g^{\mu \sigma}g^{\nu \tau}-{1 \over 2}g^{\mu \nu}g^{\sigma \tau}\right)
\nabla_{\nu} \delta^{\pm}g_{\sigma \tau}=0.
\label{(1.11)}
\end{equation}
If Eq. (1.11) is fulfilled, the Euler-Lagrange equations (1.10) take the form
\begin{equation}
\sqrt{g}\left[\left(g^{\mu \sigma}g^{\nu \tau}-{1 \over 2}g^{\mu \nu}g^{\sigma \tau}\right)g^{\rho \lambda}
\nabla_{\lambda}\nabla_{\rho}
-2 R^{\mu \sigma \nu \tau}\right]\delta^{\pm}g_{\sigma \tau}
=-2 \varepsilon {\delta B \over \delta g_{\mu \nu}}.
\label{(1.12)}
\end{equation}
One can then solve for the infinitesimal variation of the spacetime metric in the form
\begin{equation}
\delta^{\pm}g_{\mu \nu}=\varepsilon \int G_{\mu \nu \alpha \beta}^{\pm}(x,z)
{\delta B \over \delta g_{\alpha \beta}}{\rm d}^{4}z,
\label{(1.13)}
\end{equation}
where $G_{\mu \nu \alpha \beta}^{\pm}$ are the advanced and retarded Green functions for Eq.
(1.12); they satisfy
\begin{eqnarray}
\; & \; & 
g^{\sigma \tau}\nabla_{\tau}\nabla_{\sigma}G_{\mu \nu \alpha \beta}^{\pm}
-2 R_{\mu \; \; \nu}^{\; \sigma \; \; \tau} \; G_{\sigma \tau \alpha \beta}^{\pm}
\nonumber \\
&=& -\Bigr(U_{\mu \alpha}U_{\nu \beta}+U_{\mu \beta}U_{\nu \alpha}
-g_{\mu \nu}g_{\alpha \beta}\Bigr)g^{-{1 \over 4}}(x)\delta^{(4)}(x,z)g^{-{1 \over 4}}(z).
\label{(1.14)}
\end{eqnarray}
With the notation of Ref. \cite{DeWitt1960}, the indices $\alpha,\beta$ refer to the point $z$ while the
indices $\mu,\nu$ refer to the point $x$. Parallel displacement along the geodesic between $x$ and $z$ is
performed by the bivector $U_{\mu \alpha}$. The four-dimensional Dirac delta is a scalar density at both
$x$ and $z$. By construction, the Green functions $G_{\mu \nu \alpha \beta}^{\pm}$ are bi-tensors
(see the beginning of Sec. III),
by virtue of their transformation properties at two different spacetime points.
\vskip 0.3cm
\noindent
(iv) In canonical quantum gravity, inspired by the Dirac map in quantum mechanics
\cite{Dirac1932}, according to which the commutator of position and momentum operators is 
the imaginary unit times their classical Poisson bracket, DeWitt considered the commutator
defined by
\begin{equation}
[A,B] \equiv {\rm i} \int {\rm d}^{4}x \int {\rm d}^{4}z
{\delta A \over \delta g_{\mu \nu}}(x) {\widetilde {\bf G}}_{\mu \nu \alpha \beta}(x,z)
{\delta B \over \delta g_{\alpha \beta}}(z),
\label{(1.15)}
\end{equation}
where ${\widetilde {\bf G}}_{\mu \nu \alpha \beta}$ is the supercommutator for 
quantum Einstein's gravity, i.e.
\begin{equation}
{\widetilde {\bf G}}_{\mu \nu \alpha \beta} \equiv {\bf G}_{\mu \nu \alpha \beta}^{+}
-{\bf G}_{\mu \nu \alpha \beta}^{-},
\label{(1.16)}
\end{equation}
and boldface characters are here used to stress that we deal with the operator counterpart
of classical Green functions.
The work in Ref. \cite{DeWitt1960} pointed out that, if it were possible to ignore the noncommutativity
of factors on the right-hand side of Eq. (1.15), this (formal) commutator can be shown to satisfy all
identities of a quantum Poisson bracket, and hence the question of a consistent quantum theory of gravity
was reduced (but not solved!) to finding an appropriate definition of operator propagator.

We are here initiating a research program aimed at showing
that one can get pretty close to fulfilling such a task after a careful consideration 
of well-known properties of classical hyperbolic equations. For this purpose, Sec. II introduces 
fundamental solution and parametrix of a scalar wave operator; tensor wave equations are considered in
Sec. III, with emphasis on Maxwell theory in curved spacetime; nonlinear wave equations are studied
in Sec. IV, by focusing on the Choquet-Bruhat method for solving Cauchy's problem for the Einstein
equations. Sec. V proves the existence of the parametrix necessary to solve the classical Einstein
equations, by using a well-known technique for proving the existence of a parametrix for elliptic
(rather than hyperbolic!) equations. Section VI studies the role of the parametrix in the Jacobi identity
for the Peierls bracket, Sec. VII obtains novel functional equations for the parametrix, while Sec. VIII
discusses the remainder in the supercommutator function and presents our concluding remarks.
Relevant background material is described in the appendices: DeWitt's notation for gauge theories,
spacetime geometry for the wave equation, characteristic conoid and Kirchhoff formulas for the linear
hyperbolic system associated to the nonlinear Einstein equations.

\section{Fundamental solution and parametrix of a scalar wave operator}

Let us consider, for simplicity, a scalar differential operator $P$ with $C^{\infty}$ coefficients
\begin{equation}
Pu=\Box u + \langle a, \nabla u \rangle + bu=g^{\mu \nu}\nabla_{\mu} \nabla_{\nu}u
+a^{\mu}\nabla_{\mu}u+bu,
\label{(2.1)}
\end{equation}
that is defined on a connected open set $\Omega$ of four-dimensional spacetime. A {\it fundamental solution}
of $P$ is a distribution $({\cal G}_{q}(p),\phi(p))$ which is a function of the spacetime point $q$ such that
\begin{equation}
P{\cal G}_{q}-\delta_{q}=0.
\label{(2.2)}
\end{equation}
This means that \cite{Friedlander1975}
\begin{equation}
(P{\cal G}_{q},\phi)=\phi(q), \; \phi \in C_{0}^{\infty}(\Omega).
\label{(2.3)}
\end{equation}
We note that the fundamental solution is not uniquely defined, unlike the case of Green's functions
\cite{John1975}, for each of which there is a specific boundary condition.
A $C^{\infty}$ {\it parametrix} of $P$ is instead a distribution $\pi_{q}$ such that
\cite{Friedlander1975}
\begin{equation}
P {\pi}_{q}-\delta_{q} =\omega \in C^{\infty}(\Omega).
\label{(2.4)}
\end{equation}
This concept is of interest because, although a fundamental solution is a useful tool 
for solving linear partial differential equations, some
of the problems in which it plays a role can be handled better by means of functions possessing a singularity that 
is not annihilated but merely smoothed out by the differential operator under investigation. The smoothing
might even be so weak that the singularity is actually augmented but acquires a less rapid growth than was
to be anticipated from the order of the differential operator \cite{Garabedian1964}. 

If one studies the scalar wave equation on Minkowski spacetime in the coordinates
$(ct,x^{1},x^{2},x^{3})$, with $x \equiv (x^{1},x^{2},x^{3})$, the solution of the Cauchy problem
\begin{equation}
\Box \phi=0, \; \phi(t=0,x)=u_{0}(x), \;
{\partial \phi \over \partial t}(t=0,x)=u_{1}(x)
\label{(2.5)}
\end{equation}
can be expressed in the form
\begin{equation}
\phi(x,t)=\sum_{j=0}^{1}E_{j}(t)u_{j}(x),
\label{(2.6)}
\end{equation}
where, on denoting by ${\hat u}_{j}(x)$ the Fourier transforms of Cauchy data, the operators 
$E_{j}(t)$ act \cite{Treves1980}
in such a way that (here $\xi \equiv (\xi_{1},\xi_{2},\xi_{3})$)
\begin{equation}
E_{j}(t)u_{j}(x)=\sum_{k=1}^{2}(2\pi)^{-3}\int {\rm e}^{{\rm i}\varphi_{k}(x,t,\xi)}
\alpha_{jk}(x,t,\xi){\hat u}_{j}(\xi){\rm d}^{3}\xi
+R_{j}(t)u_{j}(x),
\label{(2.7)}
\end{equation}
where the $\varphi_{k}$ are real-valued {\it phase functions} which satisfy the initial condition
\begin{equation}
\varphi_{k}(t=0,x,\xi)= \langle \xi,x \rangle= x \cdot \xi
=\sum_{s=1}^{3}x^{s}\xi_{s},
\label{(2.8)}
\end{equation}
and $R_{j}(t)$ is a regularizing operator (see our earlier comments) which smoothes out the singularities
acted upon by it \cite{Treves1980}. Thus, the amplitudes $\alpha_{jk}$ and phases $\varphi_{k}$ 
make it possible to build the parametrix for the Cauchy problem. In our analysis we are going to
need the generalization of this construction to the tensor wave equation, bearing in mind that Eq. (1.14)
is the tensor version of Eq. (2.2).

Meanwhile, let us try to develop a heuristic argument on the relation between a fundamental solution ${\cal G}$ and a
$C^{\infty}$ parametrix $\pi$ of a partial differential operator $P$ as in (2.1). By omitting, for
simplicity of notation, the subscript $q$, if we consider the split
\begin{equation}
{\cal G}=\pi + Y,
\label{(2.9)}
\end{equation}
we find
\begin{equation}
P{\cal G}=P \pi + P Y 
\Longrightarrow \delta=\delta+R+PY 
\Longrightarrow Y=-P^{-1}R.
\label{(2.10)}
\end{equation}
Thus, a fundamental solution of $P$ differs from its $C^{\infty}$ parametrix by minus the inverse of $P$ 
applied to the regularizing term $R \in C^{\infty}(\Omega)$. In the simplest case, i.e. operators on the
real line, we would therefore solve the inhomogeneous equation $P \chi=\psi$ in the form
\begin{equation}
\chi(x)=\int_{y_{0}}^{{\overline y}}\pi(x,y)\psi(y){\rm d}y
-\int_{y_{0}}^{{\overline y}}\left(\int_{z_{0}}^{y}{\cal G}(x,z)R(z){\rm d}z \right)\psi(y){\rm d}y.
\label{(2.11)}
\end{equation}
Now, to the extent that the parametrix leads to a good approximate inverse of $P$, the integral
$$
\int_{y_{0}}^{{\overline y}}\pi(x,y)\psi(y){\rm d}y
$$
on the right-hand side of (2.11) provides a good approximation of 
$$
\int_{y_{0}}^{{\overline y}}{\cal G}(x,y)\psi(y){\rm d}y.
$$
Similarly, in the tensor wave equations which are the ultimate goal of our investigation, with the
associated Peierls bracket, we are going to re-express (1.1) as the sum of the parametrix and remainder
contribution, respectively.

The difference between fundamental solution and parametrix is further elucidated, with the notation
summarized in Appendix B, by sections 4.2 and 4.3 of Ref. \cite{Friedlander1975}, which prove that
there exist fundamental solutions of the operator $P$ having the form (see (B1)-(B6))
\begin{equation}
{\cal G}_{q}^{\pm}={1 \over 2\pi}\Bigr(U \delta_{\pm}(\Gamma)+V H_{\pm}(\Gamma)\Bigr),
\label{(2.12)}
\end{equation}
as well as parametrices reading as
\begin{equation}
\pi_{q}^{\pm}={1 \over 2\pi}\Bigr(U \delta_{\pm}(\Gamma)+{\widetilde V}H_{\pm}(\Gamma)\Bigr),
\label{(2.13)}
\end{equation}
where $U(p,q)$ is a smooth function belonging to $C^{\infty}(\Omega \times \Omega)$,
expressible\footnote{Following Ref. \cite{Friedlander1975}, we use local coordinates that are normal
at $q$, for which $g_{\mu \nu}(x)x^{\nu}=g_{\mu \nu}(0)x^{\nu}$, while the world function $\Gamma$ 
takes the form $\Gamma=g_{\mu \nu}(0)x^{\mu}x^{\nu}$.} by
\begin{equation}
U(x)= {\left | {g(0)\over g(x)} \right |}^{{1 \over 4}}
{\rm exp} \left(-{1 \over 2} \int_{0}^{1}a_{\mu}(rx)x^{\mu}{\rm d}r \right),
\label{(2.14)}
\end{equation}
the $a_{\mu}$ being the covariant components of the vector $a^{\mu}$ in (2.1),
while $V$ solves the characteristic initial value problem
\begin{equation}
PV=0, \; V=V_{0} \; {\rm on} \; C(q)
\label{(2.15)}
\end{equation}
and is real-analytic, hence expressible as
\begin{equation}
V=\sum_{l=0}^{\infty}V_{l}{\Gamma^{l}\over l!},
\label{(2.16)}
\end{equation}
the $V_{l}$ being elements of $C^{\infty}(\Omega \times \Omega)$ computable in the form
\begin{eqnarray}
V_{0}(x,y)&=& -{1 \over 4}U(x,y) \int_{0}^{1}
\left . {PU \over U} \right |_{x=z(s)}{\rm d}S, 
\nonumber \\
&V_{l}(x,y)=& -{1 \over 4}U(x,y) \int_{0}^{1}
\left . {PV_{l-1} \over U} \right |_{x=z(s)} s^{l} \; {\rm d}S, \;
\label{(2.17)}
\end{eqnarray}
while
\begin{equation}
{\widetilde V}=V_{0}+\sum_{l=1}^{\infty}V_{l}{\Gamma^{l}\over l!}\sigma(k_{l}\Gamma),
\label{(2.18)}
\end{equation}
where $\sigma \in C_{0}^{\infty}({\bf R})$ and takes values
\begin{equation}
\sigma(t)=1 \; {\rm if} \; |t| \leq {1 \over 2}, \;
\sigma(t)=0 \; {\rm if} \: |t| \geq 1,
\label{(2.19)}
\end{equation}
the $\left \{ k_{l} \right \}$ being a sequence of positive numbers, strictly increasing and
tending to infinity. By virtue of (2.12)-(2.18), one finds (cf. (2.10))
\begin{equation}
{\cal G}_{q}^{\pm}-\pi_{q}^{\pm}=Y_{q}^{\pm}={1 \over 2 \pi}(V-{\widetilde V})H_{\pm}(\Gamma),
\label{(2.20)}
\end{equation}
where
\begin{equation}
V-{\widetilde V}=\sum_{l=1}^{\infty}V_{l}{\Gamma^{l}\over l!}
\Bigr(1-\sigma(k_{l}\Gamma)\Bigr).
\label{(2.21)}
\end{equation}  

Note that the parametrix is not determined uniquely. If $F \in C^{\infty}(\Omega \times \Omega)$ and 
$F \sim 0$ when the world function $\Gamma \rightarrow 0$, then ${\widetilde V}+F=V_{0}$ on 
$C(q)$, the null cone of $q$, and $P({\widetilde V}+F) \sim 0$ when $\Gamma \rightarrow 0$. 
Hence one can replace ${\widetilde V}$ with ${\widetilde V}+F$ in the formulas (2.13), to 
obtain another pair of parametrices of $P$ in $\Omega$.
If $\Omega$ is a causal domain (see appendix B), the operator $P$ in (2.1) has a fundamental solution
${\cal G}_{q}^{+}$ in $\Omega$, such that
\begin{equation}
P {\cal G}_{q}^{+}=\delta_{q}, \; {\rm supp} {\cal G}_{q}^{+} \subset J^{+}(q).
\label{(2.22)}
\end{equation}
It is of the form \cite{Friedlander1975}
\begin{equation}
{\cal G}_{q}^{+}={1 \over 2 \pi}\Bigr(U \delta_{+}(\Gamma)+V^{+}\Bigr),
\label{(2.23)}
\end{equation}
with $U=U(p,q)$ defined as in (2.14), whereas, given the set
\begin{equation}
\bigtriangleup^{+} \equiv \left \{ (p,q): (p,q) \in \Omega \times \Omega, \; p \in J^{+}(q)
\right \},
\label{(2.24)}
\end{equation}
the function $V^{+}(p,q)$ is of class $C^{\infty}$ on $\bigtriangleup^{+}$ and has support
contained in $\bigtriangleup^{+}$. When the world function $\Gamma(p,q) \rightarrow 0$ for
$(p,q) \in \bigtriangleup^{+}$, the Hadamard series on the right-hand side of (2.16) is an
asymptotic expansion for $V^{+}$. In analogous way, $P$ has also a fundamental solution
${\cal G}_{q}^{-}$ in $\Omega$ such that
\begin{equation}
P {\cal G}_{q}^{-}=\delta_{q}, \; {\rm supp} {\cal G}_{q}^{-} \subset J^{-}(q),
\label{(2.25)}
\end{equation}
which is of the form
\begin{equation}
{\cal G}_{q}^{-}={1 \over 2 \pi}\Bigr(U \delta_{-}(\Gamma)+V^{-}\Bigr),
\label{(2.26)}
\end{equation}
where $V^{-}$ has properties similar to those of $V^{+}$, with past and future interchanged.

We know that there are infinitely many $C^{\infty}$ parametrices of $P$ supported in the causal
future of $q$, and the previous theorem shows that one can construct a fundamental solution 
${\cal G}_{q}^{+}$ from any one of these. However, another theorem ensures that there is in fact
{\it only one forward fundamental solution} \cite{Friedlander1975}, which is precisely the advanced
Green function needed in the construction of the Peierls bracket. Thus, Eq. (2.23) is the definitive
form of the forward fundamental solution of $P$ in $\Omega$. It may be viewed as the field of a point
source at $q$. It consists of a singular part $U \delta_{+}(\Gamma)$ which is a measure, supported
on the future null cone of $q$, and of a regular part $V^{+}$, which is a function.
The function $p \rightarrow V^{+}(p,q)$ has its support contained in $J^{+}(q)$, and lies
in $C^{\infty}(J^{+}(q))$. It is said to be the {\it tail} term of the fundamental solution; it
distinguishes the curved spacetime case from the Minkowskian case, where ${\cal G}_{q}^{+}$
is sharp, its support being the future null cone of $q$. The tail term is a solution of the
characteristic initial-value problem
\begin{equation}
PV^{+}=0, \; p \in D^{+}(q), \; V^{+}=V_{0}, \; p \in C^{+}(q).
\label{(2.27)}
\end{equation}

\section{Tensor wave equations: Maxwell theory in curved spacetime}

The tensor analogue of the differential operator $P$ in (2.1) can be written, by introducing
tensor multi-indices $I=(i_{1},...,i_{m})$ and $J=(j_{1},...,j_{m})$, in either of 
the forms \cite{Friedlander1975}
\begin{equation}
(Pu)_{I}=\nabla_{j}\nabla^{j}u_{I}+a_{I}^{\; Jj}\nabla_{j}u_{J}
+b_{I}^{\; J}u_{J},
\label{(3.1)}
\end{equation}
or
\begin{equation}
(Pu)^{I}=\nabla_{j}\nabla^{j}u^{I}+a_{\; J}^{I \; j}\nabla_{j}u^{J}
+b_{\; J}^{I}u^{J}.
\label{(3.2)}
\end{equation}
The class of $C^{\infty}$ tensor fields of rank $m$ defined on a causal domain
$\Omega_{0}$ is a vector space over the complex numbers, denoted by $E^{m}(\Omega_{0})$.
The subspace of $E^{m}(\Omega_{0})$ consisting of fields with compact support is denoted
by $D^{m}(\Omega_{0})$. A {\it tensor distribution} $u \in D^{m}(\Omega_{0})$ is then a
continuous linear map $D^{m}(\Omega_{0}) \rightarrow C$. Moreover, if $p'$ is a point of
$\Omega_{0}$, a {\it tensor-valued distribution} $T$ is a continuous linear map 
$\xi \rightarrow (T,\xi)$ of $D^{m}(\Omega_{0})$ into the finite-dimensional vector space
of tensors of rank $m$ at $p'$.

A {\it bitensor} of type $(0,m)$ at $p$ and of type $(m',0)$ at $p'$ is a multilinear form
on the product of $m$ copies of the cotangent space $T^{*}M_{p}$ and of $m'$ copies of the
tangent space $TM_{p'}$. Bitensor indices can be raised and lowered by means of the appropriate
components of the metric tensor at $p$ and at $p'$, respectively.

The Dirac distribution of rank $m$, denoted by $\delta_{p'}^{(m)}(p)$, maps 
$\xi \in D^{m}(\Omega_{0})$ to $\xi(p')$, according to
\begin{equation}
\Bigr(\delta_{p'}^{(m)}(p),\xi(p)\Bigr)=\xi(p').
\label{(3.3)}
\end{equation}
A fundamental solution of the differential operator (3.1) or (3.2) is a tensor-valued 
distribution $G_{p'}(p)$ such that
\begin{equation}
P G_{p'}(p)=\delta_{p'}^{(m)}(p).
\label{(3.4)}
\end{equation}
There exist two basic fundamental solutions $G_{p'}^{+}(p)$ and $G_{p'}^{-}(p)$ whose supports are
contained in $J^{+}(p')$ and in $J^{-}(p')$, respectively.

The {\it transport bitensor} (see $U_{\mu \alpha}$ in (1.14))
in a geodesically convex domain $\Omega$ is a bitensor field
${ }_{m}\tau(p,p')$ of rank $m$ at both $p'$ and $p$ which satisfies the differential equations
(hereafter the world function $\Gamma=\Gamma(p,p')$, and the covariant derivatives
$\nabla_{j}$ and $\nabla^{j}$ act at $p$)
\begin{equation}
\nabla^{j}\Gamma \nabla_{j}\biggr({ }_{m}\tau_{I}^{I'}\biggr)=0,
\label{(3.5)}
\end{equation}
jointly with the initial conditions
\begin{equation}
\left . \biggr({ }_{m}\tau_{I}^{I'}\biggr) \right |_{p=p'}
=\delta_{i_{1}}^{i_{1}'} ... \delta_{i_{m}}^{i_{m}'}.
\label{(3.6)}
\end{equation}
We shall also need the biscalar $\kappa(p,p')$, which solves the differential equation
\cite{Friedlander1975}
\begin{equation}
2 \langle \nabla \Gamma,\nabla \kappa \rangle +(\Box \Gamma-8)\kappa=0,
\label{(3.7)}
\end{equation}
with initial condition
\begin{equation}
\kappa(p',p')=1,
\label{(3.8)}
\end{equation}
and is given, in local coordinates, by \cite{VV1928, Morette1951}
\begin{equation}
\kappa(x,y)={\sqrt{ \left |{\rm det}{\partial^{2}\Gamma \over \partial x^{i} \partial y^{j}}
\right | } \over 4 |g(x)g(y)|^{1 \over 4}}.
\label{(3.9)}
\end{equation}
Both ${ }_{m}\tau$ and $\kappa$ are symmetric functions of their arguments. We can now state the
tensor analogue of the theorem at the end of Sec. II for the forward fundamental solution
\cite{Friedlander1975}.
\vskip 0.3cm
\noindent
{\bf Theorem} In a causal domain $\Omega_{0}$, the tensor differential operator 
\begin{equation}
(Pu)_{I}=\nabla_{j}\nabla^{j}u_{I}+b_{I}^{\; J}u_{J}
\label{(3.10)}
\end{equation}
has a fundamental solution $G_{p'}^{+}(p)$ in $\Omega_{0}$ such that
\begin{equation}
P G_{p'}^{+}(p)=\delta_{p'}^{(m)}(p), \;
{\rm supp}G_{p'}^{+}(p) \subset J^{+}(p'),
\label{(3.11)}
\end{equation}
which is of the form
\begin{equation}
G_{p'}^{+}(p)=\kappa(p,p'){ }_{m}\tau(p,p')\delta_{+}(\Gamma)+V^{+}(p,p'),
\label{(3.12)}
\end{equation}
where $V^{+}(p,p')$ is a bitensor field that vanishes if $p \not \in J^{+}(p')$ and is
of class $C^{\infty}$ on the closed set
\begin{equation}
\left \{ (p',p): (p',p) \in \Omega_{0} \times \Omega_{0}, p \in J^{+}(p')
\right \}.
\label{(3.13)}
\end{equation}

Now we remark that vacuum Maxwell theory in the absence of charges and currents is ruled precisely
by a tensor differential operator of the kind (3.10), because, upon imposing the Lorenz
\cite{Lorenz1867} supplementary (or gauge-fixing) condition, i.e.
\begin{equation}
\nabla^{\mu}A_{\mu}=0,
\label{(3.14)}
\end{equation}
the potential $A_{\mu}$ obeys the wave equation
\begin{equation}
P_{\mu}^{\; \nu}A_{\nu}=0,
\label{(3.15)}
\end{equation}
where
\begin{equation}
P_{\mu}^{\; \nu} \equiv -\delta_{\mu}^{\; \nu}\Box +R_{\mu}^{\; \nu},
\label{(3.16)}
\end{equation}
having denoted by $\Box$ the d'Alembert operator $g^{\rho \sigma}\nabla_{\rho}\nabla_{\sigma}$,
and by $R_{\mu}^{\; \nu}$ the Ricci tensor of the spacetime manifold $(M,g)$ endowed with a
Levi-Civita connection $\nabla$.

\section{Nonlinear wave equation: Einstein's equations}

So far, we have met the concepts of fundamental solution and parametrix for linear partial 
differential equations, of either scalar or tensorial nature. But how to define fundamental 
solutions for the nonlinear partial differential equations provided by Einstein's theory of
gravitation? Indeed, the vacuum Einstein equations in four spacetime dimensions are equivalent
to Ricci-flatness, i.e. $R_{\alpha \beta}=0$, where, upon defining
\begin{equation}
F^{\lambda}={1 \over \sqrt{-g}}{\partial \over \partial x^{\mu}}
\Bigr(\sqrt{-g}g^{\lambda \mu}\Bigr),
\label{(4.1)}
\end{equation}
one has the split
\begin{equation}
R_{\alpha \beta}=-L_{\alpha \beta}-N_{\alpha \beta},
\label{(4.2)}
\end{equation}
having defined \cite{Foures1952}
\begin{equation}
L_{\alpha \beta}={1 \over 2}\Bigr(g_{\alpha \mu}\partial_{\beta}+g_{\beta \mu}\partial_{\alpha}\Bigr)
F^{\mu},
\label{(4.3)}
\end{equation}
\begin{equation}
N_{\alpha \beta}={1 \over 2}g^{\lambda \mu}{\partial^{2}g_{\alpha \beta}\over 
\partial x^{\lambda} \partial x^{\mu}}+H_{\alpha \beta},
\label{(4.4)}
\end{equation}
where $H_{\alpha \beta}$ is a polynomial of covariant and contravariant metric components, and of
their first partial derivatives. Equations (4.1)-(4.3) suggest considering the supplementary
condition $F^{\lambda}=0$, so that the vacuum Einstein equations read eventually as
\begin{equation}
g^{\lambda \mu}{\partial^{2}g_{\alpha \beta}\over \partial x^{\lambda}\partial x^{\mu}}
+2 H_{\alpha \beta}=0.
\label{(4.5)}
\end{equation}
Now if we set 
\begin{equation}
g^{\lambda \mu}=A^{\lambda \mu}, \; g_{\alpha \beta}=W_{s}, \; 2H_{\alpha \beta}=f_{s},
\label{(4.6)}
\end{equation}
we realize that our gauge-fixed vacuum Einstein equations take the standard form of quasilinear
hyperbolic systems \cite{Foures1952}
\begin{equation}
A^{\lambda \mu}{\partial^{2}W_{s} \over \partial x^{\lambda} \partial x^{\mu}}
+ f_{s}=0.
\label{(4.7)}
\end{equation}
Suppose now, following again Ref. \cite{Foures1952}, that in a spacetime domain $D$, centred
at the point $\overline M$ with coordinates $(x^{i},0)$, and defined by
\begin{equation}
\left | x^{i}-{\overline x}^{i} \right | \leq d, \;
\left | x^{0} \right | \leq \varepsilon,
\label{(4.8)}
\end{equation}
and for values of the unknown functions $W_{s}$ and their first derivatives satisfying
\begin{equation}
\left | W_{s}-W_{s}({\overline M}) \right | \leq l, \;
\left | {\partial W_{s}\over \partial x^{\alpha}}
-{\partial W_{s}\over \partial x^{\alpha}}({\overline M}) \right | \leq l,
\label{(4.9)}
\end{equation}
the coefficients $A^{\lambda \mu}$ and $f_{s}$ possess partial derivatives with respect to their
arguments up to the fifth order. One can now show that, by differentiating $5$ times the Eqs. (4.7)
with respect to the variables $x^{\alpha}$, one obtains a linear system of second-order 
partial differential equations
\begin{equation}
A^{\lambda \mu}{\partial^{2}u_{s}\over \partial x^{\lambda} \partial x^{\mu}}
+B_{s}^{r \lambda}{\partial u_{r}\over \partial x^{\lambda}}+f_{s}=0,
\label{(4.10)}
\end{equation}
where the associated quadratic form $A^{\lambda \mu}X_{\lambda}X_{\mu}$ is of normal
hyperbolic type, i.e.
\begin{equation}
A^{00}>0, \; \sum_{i,j=1}^{3}A^{ij}X_{i}X_{j}<0.
\label{(4.11)}
\end{equation}
In the following calculations, it is useful to define
\begin{equation}
{\partial W_{s}\over \partial x^{\alpha}} \equiv W_{s \alpha}, \;
{\partial^{2}W_{s}\over \partial x^{\alpha} \partial x^{\beta}} 
\equiv W_{s \alpha \beta}, \; ...,
\label{(4.12)}
\end{equation}
until we denote by $U_{S}$ the partial derivatives of fifth order of $W_{s}$, i.e.
\begin{equation}
{\partial^{5}W_{s}\over \partial x^{\alpha} \partial x^{\beta}
\partial x^{\gamma} \partial x^{\delta} \partial x^{\varepsilon}}
\equiv W_{s \alpha \beta \gamma \delta \varepsilon} \equiv U_{S}.
\label{(4.13)}
\end{equation}
Differentiation of Eqs. (4.7) with respect to any variable $x^{\alpha}$ whatsoever leads to
$n$ equations having the form
\begin{eqnarray}
\; & \; & A^{\lambda \mu}{\partial^{2}W_{s \alpha}\over \partial x^{\lambda} \partial x^{\mu}}
+\left[{\partial A^{\lambda \mu}\over \partial x^{\alpha}}
+{\partial A^{\lambda \mu}\over \partial W_{r \nu}}W_{r \alpha}
+{\partial A^{\lambda \mu}\over \partial W_{r \nu}}{\partial W_{r \nu}\over \partial x^{\alpha}}
\right]{\partial W_{s \mu}\over \partial x^{\lambda}}
\nonumber \\
&+& {\partial f_{s}\over \partial x^{\alpha}}
+{\partial f_{s}\over \partial W_{r}}W_{r \alpha}
+{\partial f_{s}\over \partial W_{r \nu}}{\partial W_{r \nu}\over \partial x^{\alpha}}=0.
\label{(4.14)}
\end{eqnarray}
By repeating four times this process, the following system of $N$ equations is obtained:
\begin{eqnarray}
\; & \; & A^{\lambda \mu}{\partial^{2}W_{s \alpha \beta \gamma \delta \varepsilon}\over
\partial x^{\lambda} \partial x^{\mu}}
+\left[{\partial A^{\lambda \mu}\over \partial x^{\alpha}}
+{\partial A^{\lambda \mu}\over \partial W_{r}}W_{r \alpha}
+{\partial A^{\lambda \mu}\over \partial W_{r \nu}}W_{r \nu \alpha}\right]
{\partial \over \partial x^{\lambda}}W_{s \beta \gamma \delta \varepsilon \mu}
\nonumber \\
&+& \left[{\partial A^{\lambda \mu}\over \partial x^{\beta}}
+{\partial A^{\lambda \mu}\over \partial W_{r}}W_{r \beta}
+{\partial A^{\lambda \mu}\over \partial W_{r \nu}}W_{r \nu \beta}\right]
{\partial \over \partial x^{\lambda}}W_{s \alpha \gamma \delta \varepsilon \mu} +...
\nonumber \\ 
&+& \left[{\partial A^{\lambda \mu}\over \partial x^{\varepsilon}}
+{\partial A^{\lambda \mu}\over \partial W_{r}}W_{r \varepsilon}
+{\partial A^{\lambda \mu}\over \partial W_{r \nu}}W_{r \nu \varepsilon}\right]
{\partial \over \partial x^{\lambda}}W_{s \alpha \beta \gamma \delta \mu} 
\nonumber \\
&+& {\partial A^{\lambda \mu}\over \partial W_{r \nu}}
{\partial W_{r \nu \alpha \beta \gamma \delta}\over \partial x^{\varepsilon}}
+{\partial f_{s}\over \partial W_{r \nu}}
{\partial W_{r \nu \alpha \beta \gamma \delta}\over \partial x^{\varepsilon}}
+F_{S}=0,
\label{(4.15)}
\end{eqnarray}
where $F_{S}$ is a function of the variables $x^{\alpha}$, the unknown functions $W_{s}$ and
their partial derivatives up to the fifth order included, but not of their higher-order derivatives. 
By virtue of Eq. (4.15), the fifth derivatives $U_{S}$ of the functions $W_{s}$ satisfy, within
the domain $D$ and under the assumptions previously stated, a system of $N$ equations
\begin{equation}
A^{\lambda \mu}{\partial^{2}U_{S}\over \partial x^{\lambda} \partial x^{\mu}}
+B_{S}^{T \lambda}{\partial U_{T}\over \partial x^{\lambda}}+F_{S}=0,
\label{(4.16)}
\end{equation}
which is precisely of the linear type studied in (4.10), because
\begin{eqnarray}
\; & \; &
A^{\lambda \mu}=A^{\lambda \mu}(x^{\alpha},W_{s},W_{s \alpha}), \;
B_{S}^{T \lambda}=B_{S}^{T \lambda}(x^{\alpha},W_{s},W_{s \alpha},W_{s \alpha \beta}), 
\nonumber \\
&F_{S}&=F_{S}[x^{\alpha},W_{s},(W_{s \alpha},...,U_{S})].
\label{(4.17)}
\end{eqnarray}
By virtue of the method described in appendix C, we know that Eqs. (4.16) can be solved explicitly
through Kirchhoff-type formulas as in Eq. (C33). Thus, by virtue of (4.13), if we can find a
fundamental solution, or at least a parametrix $\pi$ of the fifth-order linear equation
\begin{equation}
{\partial^{5}\over \partial x^{\alpha} \partial x^{\beta} \partial x^{\gamma}
\partial x^{\delta} \partial x^{\varepsilon}}W_{s}=U_{S}(\{ x^{\mu} \}),
\label{(4.18)}
\end{equation}
$U_{S}$ being given by the integral formula (C33), we will know, at least in principle,
the classical Peierls bracket for gravity, which is in turn necessary to arrive at the desired
definition of quantum Peierls bracket (1.15) for gravity. We recall that the $W_{s}$ functions
in Eq. (4.18) are the covariant components of the metric tensor of the spacetime manifold. 
They can be expressed, from Eq. (4.18) and its parametrix, through the convolution formula
\begin{equation}
W_{s}=\int \pi(\{ x^{\alpha} \}, \{ y^{\beta} \}) 
U_{S}(\{ y^{\beta} \}){\rm d}\mu_{y}+{\rm remainder}.
\label{(4.19)}
\end{equation}

\section{Existence of the parametrix}

We can point out that, for every choice $({\overline \alpha},{\overline \beta},{\overline \gamma},
{\overline \delta},{\overline \varepsilon})$ of the indices $(\alpha,\beta,\gamma,\delta,\varepsilon)$,
we can set
$$
x^{\overline \alpha}=x^{1}, \; x^{\overline \beta}=x^{2}, \;
x^{\overline \gamma}=x^{3}, \; x^{\overline \delta}=x^{4}, \;
x^{\overline \varepsilon}=x^{5},
$$
and hence the operator on the left-hand side of Eq. (4.18) may be viewed as a
constant-coefficient fifth-order elliptic operator on ${\bf R}^{5}$. At this stage, we can exploit
the following theorem \cite{Hormander1983}:
\vskip 0.3cm
\noindent
{\bf Theorem}. Every elliptic operator $P(D)$ with constant coefficients has a parametrix which is
a $C^{\infty}$ function in ${\bf R}^{n}- \{ 0 \}$.
\vskip 0.3cm
\noindent
{\bf Proof}. The symbol of $P(D)$ can be written as
\begin{equation}
P(\xi)=P_{m}(\xi)+P_{m-1}(\xi)+...+P_{0},
\label{(5.1)}
\end{equation}
where the term $P_{j}$ is homogeneous of degree $j$, and $P_{m}(\xi) \not =0$ when 
$\xi \not =0$. Then there exists a constant $c$ for which
\begin{equation}
|P_{m}(\xi)| \geq c >0 
\label{(5.2)}
\end{equation}
when $| \xi |=\sqrt{(\xi_{1})^{2}+...+(\xi_{n})^{2}}=1$, 
so that the homogeneity provides the minorization
\begin{equation}
|P_{m}(\xi)|=|\xi|^{m}P_{m} \left({\xi \over |\xi|}\right) 
\geq c |\xi|^{m}, \; \xi \in {\bf R}^{n}.
\label{(5.3)}
\end{equation}
By virtue of (5.3), one can write for some constants $C$ and $R$
\begin{equation}
|P(\xi)| \geq |P_{m}(\xi)|-|P_{m-1}(\xi)|-... \geq c |\xi|^{m}
-C \biggr(|\xi|^{m-1}+...+1 \biggr) \geq c {|\xi|^{m}\over 2},
\label{(5.4)}
\end{equation}
provided that $\xi \in {\bf R}^{n}$ has $|\xi| \geq R$. Since a derivative of order $k$ of 
${1 \over P(\xi)}$ is of the form ${Q(\xi)\over P(\xi)^{k+1}}$, with $Q$ of degree 
$\leq (m-1)k$, as one can prove by induction, one finds that, when $|\xi| >R$,
\begin{equation}
\left | \xi^{\beta} D^{\alpha} \left({1 \over P(\xi)}\right) \right| \leq C_{\alpha \beta}
|\xi|^{|\beta|-|\alpha|-m}.
\label{(5.5)}
\end{equation}
Choose now $\chi \in C_{0}^{\infty}({\bf R}^{n})$ equal to $1$ in the subset of 
${\bf R}^{n}$ given by $\{ \xi: |\xi| <R \}$. This ensures that
${(1-\chi(\xi))\over P(\xi)}$ is a bounded $C^{\infty}$ function, hence it can be
viewed as the Fourier transform of a distribution $\pi$. Hence one has
\begin{equation}
P(D)\pi=\delta + \omega,
\label{(5.6)}
\end{equation}
where the Fourier transform of $\omega$ is $-\chi$. Hence $\omega$ itself is an element of
the Schwartz space, and it follows from (5.5) that $D^{\beta}x^{\alpha}\pi$ is continuous when
$|\beta|-|\alpha|-m < -n$. Q.E.D.
\vskip 0.3cm
\noindent
As is stated in Ref. \cite{Hormander1983}, the error term $\omega$ obtained in such a proof has 
an analytic extension to ${\bf C}^{n}$. Moreover, the parametrix $\pi$ admits an analytic extension
to a conic neighborhood of ${\bf R}^{n}- \{ 0 \}$. 

\section{Role of the parametrix in the Jacobi identity for the Peierls bracket}

Inspired by the split (2.9), let us first re-express the classical advanced and retarded Green functions 
mentioned in the Introduction in the form
\begin{equation}
G^{+ij}=\pi_{+}^{ij}+U_{+}^{ij},
\label{(6.1)}
\end{equation}
\begin{equation}
G^{-ij}=\pi_{-}^{ij}+U_{-}^{ij},
\label{(6.2)}
\end{equation}
where $\pi^{ij}$ is the parametrix and $U^{ij}$ is the
remainder\footnote{Strictly speaking, we should write $G^{ij'},\gamma^{ij'},R^{ij'}$, as in (1.3)
and (1.4), to stress that the indices $i,j$ refer to different spacetime points, but we refrain ourselves
from doing so for simplicity of notation.} term. Moreover, inspired by the definition (1.2), let
us introduce
\begin{equation}
{\widetilde \pi}^{ij} \equiv \pi_{+}^{ij}-\pi_{-}^{ij}, \;
{\widetilde U}^{ij} \equiv U_{+}^{ij}-U_{-}^{ij}, \; \Longrightarrow
{\widetilde G}^{ij}={\widetilde \pi}^{ij}+{\widetilde U}^{ij}.
\label{(6.3)}
\end{equation}
The lack of uniqueness of the parametrix (see comments after (2.21)) can be exploited to ensure that
${\widetilde \pi}^{ij}$ has the same antisymmetry property of the supercommutator function
${\widetilde G}^{ij}$. In other words, given the parametrices $\pi_{\pm}^{ij}$, suppose that
\begin{equation}
{\widetilde \pi}^{ij}=\pi_{+}^{ij}-\pi_{-}^{ij}={\widetilde p}^{ij}
={\widetilde p}^{(ij)}+{\widetilde p}^{[ij]},
\label{(6.4)}
\end{equation}
where ${\widetilde p}^{(ij)} \not =0$ and ${\widetilde p}^{[ij]} \not =0$. Now we can add to 
${\widetilde p}^{ij}$ a function ${\widetilde f}^{ij} \in C^{\infty}(\Omega \times \Omega)$, tending
to $0$ when $\Gamma \rightarrow 0$, such that
\begin{equation}
{\widetilde \pi}^{ij}={\widetilde p}^{ij}+{\widetilde f}^{ij} \Longrightarrow 
{\widetilde \pi}^{(ij)}={\widetilde p}^{(ij)}+{\widetilde f}^{(ij)} ,
\label{(6.5)}
\end{equation}
and hence it is enough to choose
\begin{equation}
{\widetilde f}^{(ij)}=-{\widetilde p}^{(ij)}
\label{(6.6)}
\end{equation}
to achieve the desired skew-symmetry of ${\widetilde \pi}^{ij}$. The remainder term
${\widetilde U}^{ij}$ is then skew-symmetric as well, by virtue of (6.3). 
 
In the proof that the classical Peierls bracket is a Poisson bracket on the space of gauge-invariant 
functionals, a crucial role is played by the verification of the Jacobi identity 
\cite{BEMS2003} obtained from the sum
\begin{eqnarray}
J(A,B,C)& \equiv & (A,(B,C))+(B,(C,A))+(C,(A,B))
\nonumber \\
&=& A_{,i}{\widetilde G}^{il}\Bigr(B_{,jl}{\widetilde G}^{jk}C_{,k}
+B_{,j}{\widetilde G}^{jk}C_{,kl}\Bigr)
+B_{,j}{\widetilde G}^{jl}\Bigr(C_{,kl}{\widetilde G}^{ki}A_{,i}
+C_{,k}{\widetilde G}^{ki}A_{,il}\Bigr)
\nonumber \\
&+& C_{,k}{\widetilde G}^{kl}\Bigr(A_{,il}{\widetilde G}^{ij}B_{,j}
+A_{,i}{\widetilde G}^{ij}B_{,jl}\Bigr)
\nonumber \\
&+& \Bigr[A_{,i}{\widetilde G}^{il}B_{,j}{\widetilde G}_{,l}^{jk}C_{,k}
+B_{,j}{\widetilde G}^{jl}C_{,k}{\widetilde G}_{,l}^{ki}A_{,i}
+C_{,k}{\widetilde G}^{kl}A_{,i}{\widetilde G}_{,l}^{ij}B_{,j}\Bigr],
\label{(6.7)}
\end{eqnarray}
where $J(A,B,C)$ is found to vanish classically, as shown in detail in Ref. \cite{BEMS2003} and,
first, in Ref. \cite{DeWitt1965}. By virtue of the split (6.3), the Jacobi identity (6.7) can be 
written as the sum of three terms, i.e.
\begin{equation}
J(A,B,C)=J_{\widetilde \pi}(A,B,C)+J_{{\widetilde \pi}{\widetilde U}}(A,B,C)
+J_{\widetilde U}(A,B,C),
\label{(6.8)}
\end{equation}
where $J_{\widetilde \pi}$ (respectively, $J_{\widetilde U}$) can be obtained from (6.7) 
upon replacing everywhere ${\widetilde G}^{ij}$ with ${\widetilde \pi}^{ij}$ 
(respectively, ${\widetilde U}^{ij}$), whereas $J_{{\widetilde \pi}{\widetilde U}}(A,B,C)$ denotes the 
sum of all mixed terms, e.g.
$$
A_{,i}{\widetilde \pi}^{il}(B_{,jl}{\widetilde U}^{jk}C_{,k}
+B_{,j}{\widetilde U}^{jk}C_{,kl})+... \; .
$$
Let us here focus on the term $J_{\widetilde \pi}(A,B,C)$ in the classical Jacobi identity, because,
within our scheme, it is equal to the classical part of the quantum Jacobi identity. By virtue of
(6.3) and (6.7), we find
\begin{eqnarray}
\; & \; & J_{\widetilde \pi}(A,B,C)=A_{,il}B_{,j}C_{,k}\Bigr({\widetilde \pi}^{ij}
{\widetilde \pi}^{kl}+{\widetilde \pi}^{jl}{\widetilde \pi}^{ki}\Bigr)
+A_{,i}B_{,jl}C_{,k}\Bigr({\widetilde \pi}^{jk}{\widetilde \pi}^{il}
+{\widetilde \pi}^{kl}{\widetilde \pi}^{ij}\Bigr)
\nonumber \\
&+& A_{,i}B_{,j}C_{,kl}\Bigr({\widetilde \pi}^{ki}{\widetilde \pi}^{jl}
+{\widetilde \pi}^{il}{\widetilde \pi}^{jk}\Bigr)
+A_{,i}B_{,j}C_{,k}\Bigr({\widetilde \pi}^{il}{\widetilde \pi}_{,l}^{jk}
+{\widetilde \pi}^{jl}{\widetilde \pi}_{,l}^{ki}
+{\widetilde \pi}^{kl}{\widetilde \pi}_{,l}^{ij}\Bigr).
\label{(6.9)}
\end{eqnarray}
The skew-symmetry of ${\widetilde \pi}^{ij}$, jointly with commutation of functional
derivatives: $T_{,il}=T_{,li}$ for all $T=A,B,C$, implies that the first three terms on
the right-hand side of (6.9) vanish. For example, one finds \cite{DeWitt1965,BEMS2003}
\begin{eqnarray}
\; & \; & A_{,il}B_{,j}C_{,k}\Bigr({\widetilde \pi}^{ij}{\widetilde \pi}^{kl}
+{\widetilde \pi}^{jl}{\widetilde \pi}^{ki}\Bigr)
=A_{,li}B_{,j}C_{,k}\Bigr({\widetilde \pi}^{lj}{\widetilde \pi}^{ki}
+{\widetilde \pi}^{ji}{\widetilde \pi}^{kl}\Bigr)
\nonumber \\
&=& -A_{,il}B_{,j}C_{,k}\Bigr({\widetilde \pi}^{jl}{\widetilde \pi}^{ki}
+{\widetilde \pi}^{ij}{\widetilde \pi}^{kl}\Bigr)=0,
\label{(6.10)}
\end{eqnarray}
and an identical procedure can be applied to the terms containing the second functional
derivatives $B_{,jl}$ and $C_{,kl}$ in (6.9).

\section{Functional equations for the parametrix}

The last term on the right-hand side of (6.9) requires new calculations because it contains
functional derivatives of ${\widetilde \pi}^{ij}$. These can be dealt with after taking infinitesimal
variations of an equation like (2.4). With the DeWitt notation used here, the defining equation
for the parametrix reads as
\begin{equation}
P_{ij}\pi_{\pm}^{jk}=-\delta_{i}^{k}+ \omega_{\pm \; i}^{k},
\label{(7.1)}
\end{equation}
where the signs on the right-hand side have been chosen so as to agree with Eq. (1.4) when
$\omega_{\pm}=0$ (which means that the parametrix is actually a fundamental solution or even a
Green function). The operator $P_{ij}$ is obtained, in classical theory, as \cite{DeWitt1965,BEMS2003}
\begin{equation}
P_{ij}=S_{,ij}+h_{ik}Q_{\; \alpha}^{k}l^{\alpha \beta}h_{jl}Q_{\; \beta}^{l},
\label{(7.2)}
\end{equation}
where $h_{ij}$ is a local and symmetric matrix which is taken to transform like $S_{,ij}$ under
group transformations, and $l^{\alpha \beta}$ is a local, nonsingular, symmetric matrix which
transforms according to the adjoint representation of the infinite-dimensional invariance group. 
Such matrices act as metrics that raise and lower indices of the generators $Q_{\; \alpha}^{i}$
according to the rules \cite{DeWitt1965}
\begin{equation}
Q_{i \alpha} \equiv h_{ij}Q_{\; \alpha}^{j}, \;
Q_{i}^{\; \alpha}=l^{\alpha \beta} Q_{i \beta}.
\label{(7.3)}
\end{equation}
If we study the transformation properties of Eq. (A3) under the infinitesimal gauge transformations
(A3), we find
\begin{equation}
P \pi_{\pm}=-I+\omega_{\pm} \Longrightarrow (\delta P)\pi_{\pm}+P (\delta \pi_{\pm})
=\delta \omega_{\pm}.
\label{(7.4)}
\end{equation}
Now we act with $\pi_{\pm}$ on both sides of Eq. (7.4), and we find, by virtue of (7.1), the relation
\begin{equation}
(I-\omega_{\pm})\delta \pi_{\pm}=\pi_{\pm}(\delta P)\pi_{\pm}
-\pi_{\pm} \delta \omega_{\pm}.
\label{(7.5)}
\end{equation}
At this stage we revert to condensed-index notation, hence writing, from (7.5),
\begin{equation}
\Bigr(\delta_{\; j}^{i}-\omega_{\pm \; j}^{i}\Bigr) \pi_{\pm \; ,c}^{jk}
=\pi_{\pm}^{ia} P_{ab,c} \pi_{\pm}^{bk}
-\pi_{\pm \; j}^{i} \; \omega_{\pm \; ,c}^{jk},
\label{(7.6)}
\end{equation}
where, in light of (7.2), we can write
\begin{equation}
\pi_{\pm}^{ia}P_{ab,c}\pi_{\pm}^{bk}=\pi_{\pm}^{ia}S_{,abc}\pi_{\pm}^{bk}
+\pi_{\pm}^{ia}Q_{a \alpha,c}Q_{b}^{\; \alpha} \pi_{\pm}^{bk}
+\pi_{\pm}^{ia}Q_{a \alpha} Q_{b,c}^{\; \alpha} \pi_{\pm}^{bk}.
\label{(7.7)}
\end{equation}
At this stage, we are going to re-express in (7.7) the terms $Q_{b}^{\; \alpha}\pi_{\pm}^{bk}$ and
$\pi_{\pm}^{ia}Q_{a \alpha}$. For this purpose, inspired by the method adopted in Ref. \cite{DeWitt1965},
we begin by noticing that, for background fields that solve the classical field equations, one has
(see Eq. (A7) and set $S_{,i}=0$ therein)
\begin{equation}
P_{ik}Q_{\; \alpha}^{k}=Q_{i}^{\; \rho} {\hat P}_{\rho \alpha},
\label{(7.8)}
\end{equation}
where the operator ${\hat P}_{\rho \alpha}$ is defined by \cite{DeWitt1965}
\begin{equation}
{\hat P}_{\rho \alpha} \equiv Q_{k \rho} \; Q_{\; \alpha}^{k}.
\label{(7.9)}
\end{equation}
The associated parametrix $\pi_{\pm}^{\alpha \beta}$ is here defined, by analogy with (7.1), 
in the form
\begin{equation}
{\hat P}_{\rho \alpha}\pi_{\pm}^{\alpha \beta}=-\delta_{\rho}^{\beta}
+\omega_{\pm \alpha}^{\; \; \; \beta}.
\label{(7.10)}
\end{equation}
Now we act first with $\pi_{\pm}^{ji}$ on $P_{ik}Q_{\; \alpha}^{k}$, finding, in light
of (7.1), the functional equation
\begin{equation}
Q_{\; \alpha}^{j}=\omega_{\pm \; k}^{j} \; Q_{\; \alpha}^{k}
-\pi_{\pm}^{ji} \; Q_{i}^{\; \rho} \; {\hat P}_{\rho \alpha}.
\label{(7.11)}
\end{equation}
Furthermore, we act upon Eq. (7.11) with the parametrix $\pi_{\pm}^{\alpha \beta}$ and find,
by virtue of (7.10), the functional equation resulting from (7.11), i.e.
\begin{equation}
\pi_{\pm}^{ji} Q_{i}^{\; \beta}
=Q_{\; \alpha}^{j} \; \pi_{\pm}^{\alpha \beta}
+\pi_{\pm}^{ji}Q_{i}^{\; \rho}\omega_{\pm \; \rho}^{\; \; \; \; \; \beta}
-\omega_{\pm \; k}^{j} Q_{\; \alpha}^{k} \pi_{\pm}^{\alpha \beta}.
\label{(7.12)}
\end{equation}
The left-hand side of (7.12) is exactly of the type
$\pi_{\pm}^{ia}Q_{a \alpha}$ occurring in Eq. (7.7). If we were dealing with Green functions,
Eq. (7.12) would reduce to the familiar relation \cite{DeWitt1965}
\begin{equation}
G^{\pm ji} Q_{i}^{\; \beta}
=Q_{\; \alpha}^{j} G^{\pm \alpha \beta}.
\label{(7.13)}
\end{equation}
Interestingly, on going from Green functions to parametrices, a basic functional equation
like (7.13) receives terms linear in $\omega_{\pm}^{ij}$ and $\omega_{\pm}^{\alpha \beta}$.
As a second step, we act with $\pi_{\pm}^{\beta \alpha}$ from the left on Eq. (7.8), 
finding that
\begin{equation}
Q_{i}^{\; \alpha}=Q_{i}^{\; \rho}\omega_{\pm \rho}^{\; \; \; \alpha}
-\pi_{\pm}^{\alpha \beta} P_{ik}Q_{\; \beta}^{k},
\label{(7.14)}
\end{equation}
which implies
\begin{equation}
Q_{i}^{\; \alpha} \pi_{\pm}^{ij}=\pi_{\pm}^{\alpha \beta}Q_{\; \beta}^{j}
+Q_{i}^{\; \rho}\omega_{\pm \rho}^{\; \; \; \alpha} \pi_{\pm}^{ij}
-\pi_{\pm}^{\alpha \beta}\omega_{\pm k}^{\; \; \; j} Q_{\; \beta}^{k}.
\label{(7.15)}
\end{equation}
This is the transposed of Eq. (7.12), and generalizes
the more familiar functional equation for Green functions
\cite{DeWitt1965}
\begin{equation}
Q_{i}^{\; \alpha} G^{\pm ij}
=G^{\pm \alpha \beta}Q_{\; \beta}^{j}.
\label{(7.16)}
\end{equation}
By virtue of the group invariance property satisfied by all physical observables 
(set $T=A$ or $B$ or $C$ in Eq. (A5)), the terms like
$Q_{\; \alpha}^{j} \pi_{\pm}^{\alpha \beta}$ and $\pi_{\pm}^{\alpha \beta}Q_{\; \beta}^{j}$
on the right-hand side of (7.12) and (7.15) give vanishing contribution to
$J_{\widetilde \pi}(A,B,C)$. One is therefore left with the contributions involving third
functional derivatives of the action $S$, plus terms linear in $\omega_{\pm}^{ij}$ and
$\omega_{\pm}^{\alpha \beta}$, hereafter denoted by 
${\rm O}(\omega_{\pm}^{ij},\omega_{\pm}^{\alpha \beta})$. 
Bearing in mind that $S_{,abc}=S_{,acb}=S_{,bca}=...$,one can
relabel indices summed over, finding eventually (cf. Ref. \cite{BEMS2003})
\begin{eqnarray}
J_{\widetilde \pi}(A,B,C)-{\rm O}(\omega_{\pm}^{ij},\omega_{\pm}^{\alpha \beta})
&=& A_{,i}B_{,j}C_{,k}\Bigr[(\pi_{+}^{ic}-\pi_{-}^{ic})
(\pi_{+}^{ja}\pi_{+}^{bk}-\pi_{-}^{ja}\pi_{-}^{bk})
\nonumber \\
&+& (\pi_{+}^{jc}-\pi_{-}^{jc})(\pi_{+}^{ka}\pi_{+}^{bi}
-\pi_{-}^{ka}\pi_{-}^{bi})
\nonumber \\
&+& (\pi_{+}^{kc}-\pi_{-}^{kc})(\pi_{+}^{ia}\pi_{+}^{bj}
-\pi_{-}^{ia}\pi_{-}^{bj})\Bigr]S_{,abc}
\nonumber \\
&=& A_{,i}B_{,j}C_{,k}\Bigr[(\pi_{+}^{ia}-\pi_{-}^{ia})
(\pi_{+}^{jb}\pi_{-}^{kc}-\pi_{-}^{jb}\pi_{+}^{kc})
\nonumber \\
&+& (\pi_{+}^{jb}-\pi_{-}^{jb})(\pi_{+}^{kc}\pi_{-}^{ia}
-\pi_{-}^{kc}\pi_{+}^{ia})
\nonumber \\
&+& (\pi_{+}^{kc}-\pi_{-}^{kc})(\pi_{+}^{ia}\pi_{-}^{jb}
-\pi_{-}^{ia}\pi_{+}^{jb})\Bigr]S_{,abc}=0,
\label{(7.17)}
\end{eqnarray}
where we have assumed the nontrivial properties
\begin{equation}
\pi_{+}^{ij}=\pi_{-}^{ji}, \; \pi_{-}^{ij}=\pi_{+}^{ji}.
\label{(7.18)}
\end{equation}
The sum in (7.17) vanishes because it involves six pairs of double products of parametrices with
opposite signs, i.e.
\begin{eqnarray}
\; & \; & 
\pi_{+}^{ia} \pi_{+}^{jb} \pi_{-}^{kc}-\pi_{+}^{jb}\pi_{-}^{kc}\pi_{+}^{ia}
-\pi_{+}^{ia} \pi_{-}^{jb} \pi_{+}^{kc}+\pi_{+}^{kc}\pi_{+}^{ia}\pi_{-}^{jb}
-\pi_{-}^{ia} \pi_{+}^{jb} \pi_{-}^{kc}+\pi_{-}^{kc}\pi_{-}^{ia}\pi_{+}^{jb}
\nonumber \\
&+&\pi_{-}^{ia} \pi_{-}^{jb} \pi_{+}^{kc}-\pi_{-}^{jb}\pi_{+}^{kc}\pi_{-}^{ia}
+\pi_{+}^{jb} \pi_{+}^{kc} \pi_{-}^{ia}-\pi_{+}^{kc}\pi_{-}^{ia}\pi_{+}^{jb}
\nonumber \\
&+& \pi_{-}^{jb} \pi_{-}^{kc} \pi_{+}^{ia}-\pi_{-}^{kc}\pi_{+}^{ia}\pi_{-}^{jb}=0.
\label{(7.19)}
\end{eqnarray}

\section{Remainder term in the supercommutator function and concluding remarks}

Along the years, Peierls brackets have been considered also in the modern literature to
define Poisson brackets on the space of histories \cite{Marolf94a,Marolf94b}, to
define the Feynman functional integral \cite{DeWitt2004}, to reconsider the covariant
form of Hamiltonian dynamics \cite{Ozaki2005}, to study Poisson brackets for
fermionic fields in de Sitter space \cite{EsRa2009}, and even to investigate the 
perturbative construction of models of algebraic quantum field theory \cite{Fred2015}.
The material in our Secs. II and III is necessary to achieve gradually the transition 
towards nonlinear hyperbolic equations, but at that early stage the parametrix is
neither compelling nor more useful than the standard canonical approaches. However,
since the Green functions of hyperbolic operators play such a key role in defining and
evaluating Peierls brackets, while for nonlinear hyperbolic equations not even the
fundamental solution is defined, even nowadays, our approach comes into play. We have
exploited the work in Ref. \cite{Foures1952} to argue that, since one knows how to solve
linear hyperbolic systems through integral equations, and how to obtain the solution of
nonlinear hyperbolic systems from the solution of linear ones, one can obtain good 
computational recipes for approximate evaluation of Peierls brackets, even though the exact 
Peierls bracket for gravity, obtainable in principle from the fully covariant formalism,
remains inaccessible.

More precisely, in the quantum commutator (1.15) for gravity, the factors
on the right-hand side do not commute. This means that we have to study (6.8) when
${\widetilde G}^{ij}$ is replaced by its unknown operator version, here denoted by boldface 
characters, i.e. ${\widetilde {\bf G}}^{ij}$, for which 
\begin{equation}
{\widetilde {\bf G}}^{jk}T_{,k} \not = T_{,k}{\widetilde {\bf G}}^{jk} \; \forall T=A,B,C,
\label{(8.1)}
\end{equation}
\begin{equation}
{\widetilde {\bf G}}^{il}T_{,jl} \not = T_{,jl}{\widetilde {\bf G}}^{il} \; \forall T=A,B,C.
\label{(8.2)}
\end{equation}
At this stage, our idea is to generalize the splits (6.1), (6.2) and the definitions (6.3),
{\it assuming} that the operator Green's function ${\bf G}^{ij}$ has a classical part given by
the parametrix $\pi^{ij}$ and a quantum part given by the as yet unknown operator 
${\bf U}^{ij}$, the latter being responsible for the lack of commutativity in (8.1) and (8.2).
Hence we write
\begin{equation}
{\bf G}^{ij}=\pi^{ij}+{\bf U}^{ij} \Longrightarrow 
{\widetilde {\bf G}}^{ij}={\widetilde \pi}^{ij}+{\widetilde {\bf U}}^{ij}.
\label{(8.3)}
\end{equation}
It is this operator equation that should be inserted into the operator version of
$J_{{\widetilde \pi}{\widetilde U}}(A,B,C)$ and $J_{\widetilde U}(A,B,C)$ in (6.8),
and this is the technical challenge ahead of us in the years to come.

In our paper we have developed ideas and have exploited well-established
properties of hyperbolic and elliptic partial differential equations to 
obtain a novel perspective on the commutators of canonical quantum gravity. 
We have shown that the fully diff-invariant classical Peierls bracket among diff-invariant
functionals in general relativity can be approximated by means of a parametrix of a
fifth-order elliptic operator, convoluted with functions solving Kirchhoff-type formulas.
The procedure is more systematic than the Eq. (1.14) leading to the advanced and retarded
Green functions, which was advocated but not solved in Ref. \cite{DeWitt1960}. Although the
parametrix of a partial differential operator is not unique, this apparent drawback can be 
exploited in order to evaluate with increasing accuracy the desired classical Peierls bracket.
After all, the best we can do in science is to evaluate with high accuracy the functional
relations we are interested in. We have asumed that the operator Green functions
of canonical quantum gravity consist of a classical part given by the above parametrix, plus
a quantum part that is responsible of technical complications but should be kept under control.
Our original work in Secs from V to VII puts on firm ground the existence of the desired
parametrix for gravity, and derives the novel functional equations (7.6), (7.7), (7.12) 
and (7.15) that are a concrete step towards applying parametrices in canonical quantization
of field theories possessing an infinite-dimensional invariance group.

From the point of view of general formalism, another interesting issue is whether a relation
exists with the work in Ref. \cite{Segre2005}, where the author has studied a formulation of the
ordering problems of general relativity inspired by the Gr\"{o}newald-Van Hove theorem. This 
work of Segre establishes a negative result, i.e., the lack of a suitable quantization map and of a 
suitable extension of such a map to general relativity. We hope to understand better the issue
in future, so as to investigate the possible implications (if any) for the quantum part 
${\bf U}^{ij}$ of the Green function in our Eq. (8.3). 

If it were possible to achieve our goals, a novel perspective on the longstanding problems of
(canonical) quantum gravity would emerge.

\section*{acknowledgments}
G. E. is grateful to the Dipartimento di Fisica Ettore Pancini of Federico II University for 
hospitality and support, and to B. Booss-Bavnbek for correspondence. Along the years, the author
has benefited from discussions and collaboration with G. Bimonte, G. Gionti, G. Marmo and C. Stornaiolo 
on the Hamiltonian formalism in field theory.

\begin{appendix}

\section{DeWitt notation and gauge theories}

The DeWitt notation \cite{DeWitt1965} makes it possible to use matrix-like notation for the otherwise
rather lengthy functional equations encountered in field theory. It does not solve, by itself, the open
problems of theoretical physics, but makes it possible to see easily what features are shared by 
seemingly very different field theories.

The field variables are denoted by $\varphi^{i}$, e.g. the electromagnetic potential $A_{\mu}$, 
or the Yang-Mills potential $A_{\mu}^{\alpha}$ or the spacetime metric $g_{\mu \nu}$. For
each functional $T$ of the fields, its functional derivatives are denoted by
\begin{equation}
{\delta T \over \delta \varphi^{i}}=T_{,i}=T_{1},
\label{(A1)}
\end{equation}
\begin{equation}
{\delta^{2}T \over \delta \varphi^{i}(x) \delta \varphi^{j}(x')}=T_{,ij'}=T_{2},
\label{(A2)}
\end{equation}
and so on. Since the Latin indices used here for fields carry information about the spacetime point
where the field is evaluated, an expression like $T_{,ij'}U^{j'}$ is meant to be
$$
\sum_{j} \int {\delta^{2}T \over \delta \varphi^{i}(x) \delta \varphi^{j}(x')}
U^{j}(x'){\rm d}x',
$$
unlike tensor calculus, where the Einstein convention $T_{\mu \nu}U^{\nu}$ in $n$-dimensional
spacetime is a short-hand notation for a sum without integration, i.e.
$ \sum_{\nu=0}^{n-1}T_{\mu \nu}(x)U^{\nu}(x)$.

In the gauge theories of theoretical physics, the action functional remains unchanged in value
under certain continuous changes in the dynamical variables, which are confined to finite but 
otherwise arbitrary spacetime domains. The set of all such changes constitutes a transformation group. 
The abstract group of which the transformation group forms a representation is the
{\it invariance group} of the system \cite{DeWitt1965}. Since the finite domain over which the changes
in the dynamical variables differ from zero is arbitrary, the invariance group is necessarily
infinite-dimensional, i.e. it is a pseudo-group \cite{DeWitt1965}. Under an infinitesimal group 
transformation the changes may be expressed in the general form
\begin{equation}
\delta \varphi^{i}=\int Q_{\; \alpha'}^{i} \; \delta \xi^{\alpha'}{\rm d}x'
=Q_{\; \alpha}^{i} \; \delta \xi^{\alpha},
\label{(A3)}
\end{equation}
where, for local theories, the $Q_{\; \alpha'}^{i}$ are linear combinations of Dirac's delta and its
derivatives, while the $\delta \xi^{\alpha}$ are arbitrary infinitesimal differentiable functions of the
spacetime points, known as group parameters, and vanish outside the arbitrary domain under
consideration \cite{DeWitt1965}.

When Eq. (A3) holds, the infinitesimal variation of any functional $T$ of the fields 
is expressed by
\begin{equation}
\delta T=T_{,i}\delta \varphi^{i}=T_{,i}Q_{\; \alpha}^{i} \; \delta \xi^{\alpha}.
\label{(A4)}
\end{equation}
In particular, if $T$ is gauge-invariant, its variation $\delta T$ in (A4) vanishes, and one finds
\begin{equation}
T_{,i} Q_{\; \alpha}^{i}=0 \Longrightarrow Q_{\alpha}T=0,
\label{(A5)}
\end{equation}
where $Q_{\alpha}$ are vector fields on the space of field configurations, i.e.
\begin{equation}
Q_{\alpha} \equiv Q_{\; \alpha}^{i} {\delta \over \delta \varphi^{i}}.
\label{(A6)}
\end{equation}
Note also that, if $T$ coincides with the action functional $S$ of a gauge theory, functional
differentiation of Eq. (A5) yields
\begin{equation}
S_{,ij}Q_{\; \alpha}^{i}+S_{,i}Q_{\; \alpha , j}^{i}=0.
\label{(A7)}
\end{equation}
Thus, restriction of Eq. (A7) to the dynamical subspace, where the Euler-Lagrange equations
$S_{,i}=0$ hold, tells us that the operator $S_{,ij}$ is not invertible, because it then possesses
nonvanishing eigenvectors belonging to the zero eigenvalue. Thus, one has to resort to the
imposition of suitable supplementary conditions whenever the field theory is a gauge theory
(see, for example, Eq. (1.11) taken from Ref. \cite{DeWitt1960}). For Einstein's general relativity,
the explicit form of $S_{,ij}$ was first obtained by Levi-Civita \cite{LeviCivita} and then, in
a more accessible reference, by DeWitt \cite{DeWitt1965}, but it is remarkable that, before 
any detailed calculation, the lack of Green functions or fundamental solutions of $S_{,ij}$ 
for any gauge theory is already clear from first principles.

\section{Spacetime geometry for the wave equation}

A connected open set $D$ of the spacetime manifold $(M,g)$ is said to be a geodesically
convex domain if any two points $q$ and $p$ in $D$ are joined by a unique geodesic in $D$. In a
geodesically convex domain $\Omega$, let the local coordinates of two points $p$ and $q$ be
$x$ and $y$, respectively. If $\rho: \tau \rightarrow z(\tau), \tau \in [0,1]$, is a parametrized
$C^{2}$ curve joining $q$ and $p$ in $\Omega$, the arc-length of $\rho$ is defined by
\begin{equation}
s \equiv \int_{0}^{1}\sqrt{g_{\mu \nu}(z(\tau)){{\rm d}z^{\mu}\over {\rm d}\tau}
{{\rm d}z^{\nu}\over {\rm d}\tau}}\; {\rm d}\tau .
\label{(B1)}
\end{equation}
When $\rho$ is the unique geodesic joining $q$ and $p$ in $\Omega$, the arc-length $s$ is said
to be the {\it geodesic distance} of $q$ and $p$, and the square of the geodesic distance is
$\Gamma(p,q)$: 
\begin{equation}
\Gamma(p,q) \equiv s^{2},
\label{(B2)}
\end{equation}
and is written as $\Gamma(x,y)$ in local coordinates. In the literature on general relativity, 
$\Gamma(p,q)$ is called the world function. It was first introduced by Hadamard 
\cite{Hadamard} in the analysis
of the Cauchy problem for linear partial differential equations, 
then by Ruse \cite{Ruse1931} and Synge \cite{Synge1931}, and 
eventually by DeWitt and Brehme \cite{DeWittBrehme1960} in the 
investigation of electromagnetic Green functions on curved spacetime. 

In a geodesically convex domain $\Omega$, the future (resp. past) null semi-cone
$C^{+}(q)$ (resp. $C^{-}(q)$) is the set of all points $p$ of $\Omega$ such that there exists a
future-directed (resp. past-directed) null geodesic from $q$ to $p$. The future (resp. past) domain of 
dependence $D^{+}(q)$ (resp. $D^{-}(q)$) 
is the set of all points $p \in \Omega$ that can be reached along 
future-directed (resp. past-directed) timelike geodesics from $q$. One has therefore
\begin{equation}
C^{+}(q)=\partial D^{+}(q), \; C^{-}(q)=\partial D^{-}(q).
\label{(B3)}
\end{equation}
The closure of $D^{+}(q)$ (resp. $D^{-}(q)$) is instead called the future 
(resp. past) emission of $q$:
\begin{equation}
J^{+}(q) \equiv {\overline {D^{+}(q)}}, \;
J^{-}(q) \equiv {\overline {D^{-}(q)}}.
\label{(B4)}
\end{equation}
Such sets are also called causal future ($J^{+}$) and causal past ($J^{-}$) of $q$
\cite{HawkingEllis1973}.

In Eqs. (2.12) and (2.13), we have 
\begin{equation}
H_{\pm}(\Gamma) \equiv 1 \; {\rm if} \; p \in J^{\pm}(q), \;
0 \; {\rm if} \; p \not \in J^{\pm}(q).
\label{(B5)}
\end{equation}
Moreover, 
\begin{equation}
U \delta_{\pm}(\Gamma) \equiv U(p,q) \delta_{\pm}(\Gamma(p,q))
\label{(B6)}
\end{equation}
are distributions that act on test functions $\phi(p) \in C_{0}^{\infty}(\Omega)$.

A connected open set $\Omega$ is called a {\it causal domain} if
\vskip 0.3cm
\noindent
(i) there is a geodesically convex domain $\Omega_{0}$ such that $\Omega \subset \Omega_{0}$, and
\vskip 0.3cm
\noindent
(ii) $\forall p,q \in \Omega$, the set $J^{+}(q) \cap J^{-}(p)$ is either a compact subset of
$\Omega$, or otherwise is empty.

\section{Characteristic conoid and Kirchhoff formulas}

For systems of linear second-order partial differential equations, the theory of characteristics 
is developed starting from the equations \cite{LeviCivita1931}
\begin{equation}
E_{s}=\sum_{r=1}^{m}\sum_{\lambda,\mu=0}^{n}E_{\; sr}^{\lambda \mu}
{\partial^{2}\varphi_{r}\over \partial x^{\lambda} \partial x^{\mu}}+\Phi_{s}
=\sum_{r=1}^{m}E_{\; sr}^{00}{\partial^{2}\varphi_{r}\over \partial (x^{0})^{2}}+...=0,
\label{(C1)}
\end{equation}
which are soluble with respect to ${\partial^{2} \varphi_{r} \over \partial (x^{0})^{2}}$ 
(this is the {\it normality} condition) if
$\Omega \equiv {\rm det}E_{\; sr}^{00} \not=0$. We then ask ourselves under which conditions
is the normality character preserved if, to the independent variables 
$x^{0},x^{1},...,x^{n}$, we apply the transformation
$$
(x^{0},x^{1},...,x^{n}) \rightarrow (z,z^{1},...,z^{n})
$$
so that the hyperplane $x^{0}=a^{0}$ is transformed into an hypersurface having equation
\begin{equation}
z(x^{0},x^{1},...,x^{n})=z^{0},
\label{(C2)}
\end{equation}
starting from which one can determine, at least in a neighborhood, the $\varphi_{r}$ functions.
For this purpose, let us consider the covector (or covariant vector or one-form) variables
\begin{equation}
\xi_{\mu} \equiv {\partial z \over \partial x^{\mu}},
\label{(C3)}
\end{equation}
from which we get \cite{LeviCivita1931}
\begin{equation}
{\partial \varphi_{r}\over \partial x^{\mu}}
={\partial \varphi_{r}\over \partial z}\xi_{\mu}
+\sum_{\lambda=1}^{n}{\partial \varphi_{r}\over \partial z^{\lambda}}{\partial z^{\lambda}\over \partial x^{\mu}}
={\partial \varphi_{r}\over \partial z}\xi_{\mu}+...,
\label{(C4)}
\end{equation}
\begin{equation}
{\partial^{2}\varphi_{r}\over \partial x^{\lambda} \partial x^{\mu}}
={\partial^{2}\varphi_{r}\over \partial z^{2}}\xi_{\lambda}\xi_{\mu}+... \; .
\label{(C5)}
\end{equation}
Hence the original system (C1) gets transformed into 
\begin{equation}
\sum_{r=1}^{m}{\partial^{2}\varphi_{r} \over \partial z^{2}}
\sum_{\lambda,\mu=0}^{n}E_{\; sr}^{\lambda \mu} \xi_{\lambda}\xi_{\mu}+...=0, \;
s=1,2,...,m.
\label{(C6)}
\end{equation}
Thus, upon considering the principal (or leading) symbol of the operator in Eqs. (C1), i.e.
the matrix of polynomials
\begin{equation}
\omega_{sr} \equiv \sum_{\lambda,\mu=0}^{n}E_{\; sr}^{\lambda \mu}\xi_{\lambda}\xi_{\mu},
\label{(C7)}
\end{equation}
also called the characteristic polynomial, the normality condition of (C1) is preserved
if $\Omega \equiv {\rm det}\omega_{sr} \not =0$. If instead $\Omega$ vanishes, it is no
longer possible to apply (whatever the value of $z^{0}$) the Cauchy theorem starting from the
hypersurfaces $z=z^{0}$. The equation 
\begin{equation}
\Omega={\rm det}\omega_{sr}=0
\label{(C8)}
\end{equation}
is the equation defining the {\it characteristics manifolds} \cite{LeviCivita1931}.

In particular, the characteristic surfaces of the system (4.10) are three-dimensional manifolds 
(null hypersurfaces \cite{Friedlander1975}) of four-dimensional spacetime with coordinates
$x^{\alpha}$ and solve the differential system
\begin{equation}
F=\sum_{\lambda,\mu=0}^{3}A^{\lambda \mu}y_{\lambda}y_{\mu}=0,
\label{(C9)}
\end{equation}
\begin{equation}
\sum_{\lambda=0}^{3}y_{\lambda}{\rm d}x^{\lambda}=0.
\label{(C10)}
\end{equation}
The four quantities $y_{\lambda}$ denote a system of directional parameters of the normal vector
field. Let us take this system, which is only defined up to a proportionality factor, in such a way
that $y_{0}=1$, and let us set $y_{i}=\xi_{i}$. The desired surfaces are therefore a solution of
\begin{equation}
F=A^{00}+2\sum_{i=1}^{3}A^{i0}\xi_{i}+\sum_{i,j=1}^{3}A^{ij}\xi_{i}\xi_{j}=0,
\label{(C11)}
\end{equation}
\begin{equation}
{\rm d}x^{0}+\sum_{i=1}^{3}\xi_{i}{\rm d}x^{i}=0.
\label{(C12)}
\end{equation}
The characteristics of this differential system, which are bicharacteristics (i.e. null 
geodesics \cite{Friedlander1975}) of the Eqs. (4.10), satisfy the following differential equations:
\begin{equation}
{{\rm d}x^{i}\over \left(A^{i0}+\sum_{j=1}^{3}A^{ij}\xi_{j}\right)}
={{\rm d}x^{0}\over \left(A^{00}+\sum_{i=1}^{3}A^{i0}\xi_{i}\right)}
=-{{\rm d}\xi_{i}\over {1 \over 2}\left({\partial F \over \partial x^{i}}
-\xi_{i}{\partial F \over \partial x^{0}}\right)}
={\rm d}\lambda_{1},
\label{(C13)}
\end{equation}
$\lambda_{1}$ being an auxiliary parameter. The characteristic conoid $\Sigma_{0}$ with vertex
$M_{0}(x_{0}^{\alpha})$ is the characteristic surface generated from the bicharacteristics passing
through $M_{0}$. Such a bicharacteristic solves the system of integral equations \cite{Foures1952}
\begin{equation}
x^{i}=x_{0}^{i}+\int_{0}^{\lambda_{1}}T^{i} \; {\rm d}\lambda_{1}, \;
T^{i} \equiv A^{i0}+A^{ij}\xi_{j},
\label{(C14)}
\end{equation}
\begin{equation}
x^{0}=(x_{0})^{0}+\int_{0}^{\lambda_{1}}T^{0} \; {\rm d}\lambda_{1}, \;
T^{0} \equiv A^{00}+\sum_{i=1}^{3}A^{i0}\xi_{i},
\label{(C15)}
\end{equation}
\begin{equation}
\xi_{i}=\xi_{i}^{0}+\int_{0}^{\lambda_{1}}R_{i} \; {\rm d}\lambda_{1}, \;
R_{i} \equiv -{1 \over 2}\left({\partial F \over \partial x^{i}}
-\xi_{i}{\partial F \over \partial x^{0}}\right),
\label{(C16)}
\end{equation}
where the $\xi_{i}^{0}$ satisfy the relation (C3), i.e.
\begin{equation}
A_{0}^{00}+2\sum_{i=1}^{3}A_{0}^{i0}\xi_{i}^{0}+\sum_{i,j=1}^{3}A_{0}^{ij}\xi_{i}^{0}\xi_{j}^{0}=0,
\label{(C17)}
\end{equation}
$A_{0}^{\lambda \mu}$ being the value of $A^{\lambda \mu}$ at the vertex $M_{0}$ of the
conoid $\Sigma_{0}$. One can assume the following values:
\begin{equation}
A_{0}^{00}=1, \; A^{i0}=0, \; A_{0}^{ij}=-\delta^{ij},
\label{(C18)}
\end{equation}
so that (C17) reduces to
\begin{equation}
\sum_{i=1}^{3}(\xi_{i}^{0})^{2}=1.
\label{(C19)}
\end{equation}
Besides the parameter $\lambda_{1}$ which defines the position of a point on a given 
bicharacteristic, we will need two more parameters $\lambda_{2}$ and $\lambda_{3}$ which vary
with the bicharacteristic under investigation and are given by \cite{Foures1952}
\begin{equation}
\xi_{1}^{0}=(\sin \lambda_{2})(\cos \lambda_{3}), \;
\xi_{2}^{0}=(\sin \lambda_{2})(\sin \lambda_{3}), \;
\xi_{3}^{0}=\cos \lambda_{2}.
\label{(C20)}
\end{equation}
Moreover, following again Ref. \cite{Foures1952}, for any function $\varphi$ of the 
spacetime coordinates $x^{\alpha}$, we denote by $[\varphi]$ its restriction to the
characteristic conoid, i.e.
\begin{equation}
[\varphi] \equiv \left. \varphi(x^{\alpha}) \right |_{\Sigma_{0}},
\label{(C21)}
\end{equation}
and we consider the second-order operator 
\begin{equation}
M(\varphi) \equiv \sum_{i,j=1}^{3} [A^{ij}]
{\partial^{2}\varphi \over \partial x^{i} \partial x^{j}},
\label{(C22)}
\end{equation}
and its (formal) adjoint
\begin{equation}
{\overline M}(\psi) \equiv \sum_{i,j=1}^{3}{\partial^{2}([A^{ij}]\psi) \over \partial x^{i}
\partial x^{j}}.
\label{(C23)}
\end{equation}
At this stage, the work in Ref. \cite{Foures1952} took linear combinations of Eqs. (4.10),
with left-hand side (with free index $r$) denoted by $E_{r}$, the coefficients of linear combination
being some a priori unknown functions $\sigma_{s}^{r}$. Remarkably, such functions were found to solve
a set of equations and turned out to obey the factorization property
\begin{equation}
\sigma_{s}^{r}= \theta \; \omega_{s}^{r},
\label{(C24)}
\end{equation}
where
\begin{equation}
\omega_{s}^{r}=\int_{0}^{\lambda_{1}}\Bigr(\sum_{t=1}^{3}Q_{t}^{r} \omega_{s}^{t}
+Q \omega_{s}^{r}){\rm d}\lambda_{1}+\delta_{s}^{r},
\label{(C25)}
\end{equation}
having defined
\begin{equation}
Q_{t}^{r} \equiv {1 \over 2} \left([B_{t}^{r0}]+[B_{t}^{ri}]\xi_{i}\right),
\label{(C26)}
\end{equation}
\begin{equation}
Q \equiv -{1 \over 2}\sum_{i,j=1}^{3} \left(\xi_{j}{\partial \over \partial x^{i}}[A^{ij}]\right)
-{1\over 2}\sum_{i=1}^{3}\left({\partial \over \partial x^{i}}[A^{i0}]\right),
\label{(C27)}
\end{equation}
while, on denoting by
\begin{equation}
J \equiv {D(x^{1},x^{2},x^{3})\over D(\lambda_{1},\lambda_{2},\lambda_{3})}
\label{(C28)}
\end{equation}
the Jacobian of the change of variables $x^{i}=x^{i}(\lambda_{j})$ on the conoid $\Sigma_{0}$,
one finds \cite{Foures1952}
\begin{equation}
\theta={\sqrt{|\sin \lambda_{2}|}\over \sqrt{|J|}}, \;
\lim_{\lambda_{1} \to 0}\theta \lambda_{1}=1.
\label{(C29)}
\end{equation}
The work in Ref. \cite{Foures1952} arrived at Kirchhoff formulas for the solution of Eqs. (4.10) 
where the integrand is built from the functions
\begin{eqnarray}
E_{s}^{i}&=& \sum_{j,r=1}^{3} \biggr \{ 
[A^{ij}]\sigma_{s}^{r}{\partial [u_{r}]\over \partial x^{j}}
-[u_{r}]{\partial \over \partial x^{j}}([A^{ij}]\sigma_{s}^{r})
+[B_{ri}^{t}][u_{t}]\sigma_{s}^{r} 
\nonumber \\ 
&+& 2 \sigma_{s}^{r}\left \{ [A^{ij}]\xi_{j}+[A^{i0}] \right \}
\left[{\partial u_{r}\over \partial x^{0}}\right] \biggr \},
\label{(C30)}
\end{eqnarray}
\begin{equation}
L_{s}^{r}={\overline M}(\sigma_{s}^{r})-\sum_{i=1}^{3} {\partial \over \partial x^{i}}
\left([B_{t}^{ri}]\sigma_{s}^{t}\right),
\label{(C31)}
\end{equation}
and also from the parameter $T$ resulting from the following geometric considerations.

The surfaces upon which we perform integration are surfaces $x^{0}={\rm constant}$ traced on the
characteristic conoid $\Sigma_{0}$. Thus, in light of (C10), they fulfill the differential
relation $p_{i}{\rm d}x^{i}=0$. To compute the surface element ${\rm d}S$ we re-express the 
volume element, originally written as
$$
{\rm d}V=\prod_{i=1}^{3}{\rm d}x^{i}=\prod_{i=1}^{3}{\rm d}\lambda_{i},
$$
by exploiting the surfaces $x^{0}={\rm constant}$ and the bicharacteristics (where only
$\lambda_{1}$ is varying), i.e. \cite{Foures1952}
\begin{equation}
{\rm d}V=(\cos \nu) \sqrt{|T|} \; {\rm d}\lambda_{1} \; {\rm d}S,
\label{(C32)}
\end{equation}
where $\sqrt{|T|} \; {\rm d}\lambda_{1}$ is the length element of the bicharacteristic, and $\nu$ is
the angle of the bicharacteristic with the normal to the surface $S$ at the point considered.

Having defined all the concepts we need, we can now state the theorem proved in Ref. \cite{Foures1952}
for linear hyperbolic systems.
\vskip 0.3cm
\noindent
{\bf Theorem}. Let the linear hyperbolic system (4.10) be given, satisfying the following assumptions:
\vskip 0.3cm
\noindent
(i) At the point $M_{0}$, the conditions (C18) hold.
\vskip 0.3cm
\noindent
(ii) The functions $A^{\lambda \mu}$ and $B_{s}^{r \lambda}$ have partial derivatives with respect
to $x^{\alpha}$ of order $4$ and $2$, respectively, continuous and bounded in a domain
$$
D: \left |x^{i}-{\overline x}^{i} \right | \leq d, \;
\left |x^{0} \right | \leq \varepsilon.
$$
The functions $f_{r}$ are continuous and bounded.
\vskip 0.3cm
\noindent
(iii) The fourth partial derivatives of $A^{\lambda \mu}$ and the second partial derivatives
of $B_{s}^{r \lambda}$ fulfill Lipschitz conditions.
\vskip 0.3cm
\noindent
Then every continuous, bounded solution of the Eqs. (4.10) with continuous and bounded first derivatives within $D$ 
satisfies the Kirchhoff integral relations
\begin{eqnarray}
4 \pi u_{s}(x_{j})&=& \sum_{r=1}^{3} \biggr[\int_{(x_{0})^{0}}^{0} \int_{0}^{2 \pi} \int_{0}^{\pi}
\left([u_{r}]L_{s}^{r}+\sigma_{s}^{r}[f_{r}]\right) {J \over T^{0}}
{\rm d}x^{0} \; {\rm d}\lambda_{2} \; {\rm d}\lambda_{3}
\nonumber \\ 
&+& \int_{0}^{2 \pi} \int_{0}^{\pi}
\left \{{E_{s}^{r}J \xi_{r}\over T^{0}} \right \}_{x^{0}=0}
{\rm d}\lambda_{2} \; {\rm d}\lambda_{3} \biggr],
\label{(C33)}
\end{eqnarray}
if the coordinates $x_{0}^{\alpha}$ of $M_{0}$ fulfill majorizations of the form
\begin{equation}
\left | (x_{0})^{0} \right | \leq \varepsilon_{0}, \;
\left | x_{0}^{i}-{\overline x}^{i} \right | \leq d,
\label{(C34)}
\end{equation}
which define a domain $D_{0} \subset D$.

\end{appendix}

\end{document}